**Spin absorption by *in situ* deposited nanoscale magnets on graphene spin valves**

Walid Amamou[1,*], Gordon Stecklein[2,*], Steven J. Koester[3], Paul A. Crowell[2], and Roland K. Kawakami[1,4]

1) Program of Materials Science and Engineering, University of California, Riverside, California 92521, United States
2) School of Physics and Astronomy, University of Minnesota, Minneapolis, Minnesota 55455, United States
3) Department of Electrical and Computer Engineering, University of Minnesota, Minneapolis, Minnesota 55455, United States
4) Department of Physics, The Ohio State University, Columbus, Ohio 43210, United States
*W.A. and G.S. contributed equally to this work.

An *in situ* measurement of spin transport in a graphene nonlocal spin valve is used to quantify the spin current absorbed by a small (250 nm x 750 nm) metallic island. The experiment allows for successive depositions of either Fe or Cu without breaking vacuum, so that the thickness of the island is the only parameter that is varied. Furthermore, by measuring the effect of the island using separate contacts for injection and detection, we isolate the effect of spin absorption from any change in the spin injection and detection mechanisms. As inferred from the thickness dependence, the effective spin current $j_e = \frac{2e}{\hbar} j_s$ absorbed by Fe is as large as $10^8$ A/m². The maximum value of $j_e$ is limited by the resistance-area product of the graphene/Fe interface, which is as small as 3 Ωμm². The spin current absorbed by the same thickness of Cu is smaller than for Fe, as expected given the longer spin diffusion length and larger spin resistance of Cu compared to Fe. These results allow for a quantitative assessment of the prospects for achieving spin transfer torque switching of a nanomagnet using a graphene-based nonlocal spin valve.

Graphene is a promising material for lateral spin transport due to its low spin orbit coupling and high carrier mobility, leading to long spin diffusion lengths at room temperature[1,2]. The graphene/ferromagnet (FM) interface has proven to be the bottleneck for achieving high spin lifetimes and high spin injection efficiencies, due to spin absorption by the ferromagnetic contacts[3-13], the possibility of contact-induced spin relaxation mechanisms other than spin absorption[14], and the challenge of separating these effects from the spin injection and detection efficiencies of the ferromagnet contacts. Understanding spin relaxation and spin absorption at graphene/FM junctions is also important for technological applications such as all-spin logic, in which the magnetization of a nanomagnet is switched by spin-transfer torque when a pure spin current is absorbed[15]. Despite apparent progress[16-17], this goal has been challenging to achieve in graphene, which is why an experiment observing the evolution of the absorbed spin current while varying the thickness of the nanomagnet is valuable.

In this Letter, we quantify the spin current absorbed by a nanomagnetic island deposited on a nonlocal graphene spin valve. Spin transport measurements are completed *in situ* while growing the Fe island and the results are interpreted using a 2-D finite-element model. We determine that the effective spin current absorbed by Fe is as large as $j_e = \frac{2e}{\hbar} j_s = 10^8$ A/m² and is limited by the resistance-area product of the graphene/Fe interface, which we find to be as low as 3 $\Omega\mu$m². When the Fe is replaced with Cu, the spin current absorbed by the same thickness is smaller than for Fe, as expected given the longer spin diffusion length and larger spin resistance of Cu compared to Fe.

The experiments were conducted in an ultrahigh vacuum chamber with the ability to perform *in situ* spin transport measurements between sequential depositions of metallic adatoms. The experiments used multilayer graphene (MLG) spin valves held at cryogenic temperatures (~20 K). For the fabrication of spin valve devices, MLG flakes were exfoliated onto SiO$_2$(300 nm)/Si, where the degenerately doped Si was used as a back gate. A single e-beam lithography pattern of bilayer PMMA/MMA resist combined with multi-angle shadow evaporation was utilized to define an SrO tunnel barrier and 60 nm thick Co electrodes. Details of the device fabrication are provided elsewhere[18]. Subsequently, the spin valve device was spin coated with PMMA and soft baked at 50 °C for 2h in order to prevent damaging the SrO/Co electrodes. After the bake, an opening in the PMMA was created for the island deposition using e-beam lithography for patterning and developing in MIBK/IPA for 1m 20s. Immediately after development, the sample was loaded into the chamber to perform the nonlocal spin transport measurement and deposit the Fe or Cu island, which was carried out by cryogenic MBE at a pressure below 1 x 10$^{-10}$ Torr. Unless otherwise noted, the samples were held at 20 K during growth in order to reduce the lateral surface diffusion of adatoms on graphene[19]. A schematic of the device and an SEM image after measurement are shown in Fig. 1(a) and 1(b). Thermal effusion cells were utilized for the island growth and the growth rate was measured using a quartz crystal monitor. Typical growth rates were 0.04 Å/min for Fe and 0.2 Å/min for Cu.

The experiment was performed four times, which we refer to as Experiments 1-4. Spin transport measurements of the MLG spin valve devices were carried out at 20 K using lock-in detection with an AC injection current of $I = 1$ $\mu$A rms at 11 Hz. Initial measurements were conducted at gate voltage $V_g = 0$ V. Injected spins diffuse throughout the graphene channel and generate the nonlocal voltage $V_{NL}$ indicated in Fig. 1(a). Spin transport was detected by monitoring the nonlocal resistance $R_{NL}$, defined as $R_{NL} = V_{NL}/I$, as a function of in-plane magnetic field applied along the long axes of the FM electrodes. Fig. 1(c) shows data for a representative MLG spin valve (Experiment 1) with MLG of width 2.5 $\mu$m and injector-detector electrode spacing of 1.5 $\mu$m. The abrupt change in $R_{NL}$, labeled as $\Delta R_{NL}$, occurred when the injector and detector magnetizations switched between parallel and antiparallel alignment. Notably, $\Delta R_{NL}$ quantifies the spin accumulation at the detector. To determine the spin lifetime and diffusion constant for the MLG, we performed nonlocal Hanle spin precession measurements in which an out-of-plane magnetic field was ramped while measuring $R_{NL}$. The Hanle curves for parallel and antiparallel magnetization states are shown in Fig. 1(d). Following a procedure described elsewhere[20], we fit the difference of the Hanle curves and obtained a spin lifetime of $\tau_s$ = 585 ps, a diffusion constant of $D$ = 35 cm$^2$/s, and spin injection/detection efficiency $\alpha$ = 11.3% (see Supplemental Material for a comparison of the data and fit). This fit accounts for spin absorption by the graphene/SrO/Co contacts according to the measured contact resistances, which were typically 2-5 k$\Omega$. The spin transport properties of the graphene channel were assumed to be constant as the island is deposited.

To investigate the effect of depositing Fe, we measured the spin transport signal $\Delta R_{NL}$ as a function of Fe thickness ($t_{Fe}$). After each cycle of Fe deposition, we slightly adjusted $V_g$ to maintain a constant channel resistance to offset any electrical doping or current shunting effects. Fig. 2(a) shows representative nonlocal magnetoresistance scans at several thicknesses, and Fig. 2(b) summarizes $\Delta R_{NL}(t_{Fe})$. The most important feature of the data is the reduction of $\Delta R_{NL}$ as Fe thickness increases. Most of the signal reduction occurred within the first monolayer of deposition, which can be most easily seen in the inset of Fig. 2(b), which shows the Fe thickness on a logarithmic scale. The thickness of a monolayer of bcc Fe is $t_{ML}$ = 1.43 Å for (001) planes[21].

The measured thickness dependence of $\Delta R_{NL}$ is interpreted as follows. We quantify the decrease in $\Delta R_{NL}$ in terms of the effective increase $\Gamma$ in the spin relaxation rate under the island. As the island is deposited, we assume the local spin relaxation effectively increases such that

$$\tau_s^{-1} \rightarrow \tau_s^{-1} + \Gamma. \tag{1}$$

This effect is represented schematically in Fig. 3(a).

Next, we introduce a theoretical model for the thickness dependence of $\Gamma$ based on spin absorption. For convenience, we convert spin current $j_s$, which has units of spin angular momentum per unit area per unit time, to effective spin current $j_e = \frac{2e}{\hbar} j_s$. Due to spin absorption, the thickness dependence of $\Gamma(t)$ is[22,23]

$$\Gamma(t) = \frac{R_{sq}D}{\frac{\rho\lambda}{\tanh(t/\lambda)} + \frac{R_I A}{\tanh(t/t_{ML})}}, \tag{2}$$

where $\rho$ is the resistivity of the island, $\lambda$ is the spin diffusion length of the island, and $R_I A$ is the resistance-area product of the graphene/island interface, which we assume fully forms when the island reaches thickness $t_{ML}$. The derivation of Eq. (2) is discussed in the Supplemental Material. The relationship between thickness and $\Gamma$ corresponding to Experiments 1 (Fe) and 4 (Cu) is shown in Fig. 3(b) for various values of $R_I A$. For Fe we assume $t_{ML}$ = 1.43 Å, $\rho_{Fe}$(20 K) = 13 µΩcm, and $\lambda_{Fe}$ = 40 Å[24], for which the spin resistance of the fully formed Fe island is $\rho_{Fe}\lambda_{Fe}$ = 0.5 fΩm², as consistent with results reported elsewhere[25]. We note that without any series resistance, $\Gamma$ reaches large values (over 1000 ns⁻¹) at very small thicknesses ($t_{Fe}$ < 0.01 Å) because of the large mismatch between the graphene and Fe spin resistances, which shows that the interface resistance must be taken into account. For Cu we assume $t_{ML}$ = 1.81 Å, $\rho_{Cu}$(20 K) = 10 µΩcm (for ultrathin films)[26], and $\lambda_{Cu}$ = 6000 Å[27], for which $\rho_{Cu}\lambda_{Cu}$ = 60 fΩm². As shown in Fig. 3(b), the effect of the Cu island is observed at larger thicknesses than for Fe because of the longer spin diffusion length and larger spin resistance of Cu compared to Fe.

To determine the sensitivity of $\Delta R_{NL}$ to $\Gamma$, we perform a 2D finite-element simulation to model the spin accumulation throughout the graphene channel, as indicated in Fig. 3(a)[28]. Details of the model are provided in the Supplemental Material. With this model, we determine the sensitivity curve $\Delta R_{NL}(\Gamma)$, which describes how the spin accumulation under the detector is affected by additional spin relaxation rate under the island, including the effect of spins flowing through side channels around the island. A plot of $\Delta R_{NL}(\Gamma)$ for Experiment 1 is shown in Fig. 3(c). We find that variation in $\Delta R_{NL}$ occurs when the additional spin relaxation rate is in the range $\Gamma = 1 - 1000$ ns$^{-1}$. Any further increase in $\Gamma$ has little effect on $\Delta R_{NL}$ because the effective spin lifetime under the island is already negligibly small.

The theoretical thickness dependence $\Gamma(t_{Fe})$ shown in Fig. 3(b) is combined with the sensitivity curve $\Delta R_{NL}(\Gamma)$ shown in Fig. 3(c) to determine $\Delta R_{NL}(t_{Fe})$ for a given value of $R_I A$. We fit the measured thickness dependence $\Delta R_{NL}(t_{Fe})$ to the spin absorption model using $R_I A$ as the sole fitting parameter. The data and fits for Experiments 1-3 are shown in Fig. 4(a), where the experiments encompass various geometries, Fe deposition temperatures, and spin transport properties. In all cases we are able to fit the data to the spin absorption model assuming $\lambda_{Fe} = 4$ nm and $t_{ML} = 1.43$ Å.

Using the best-fit value of $R_I A$, the spin current absorbed by the island in the limit $t \gg \lambda$ is calculated by

$$j_e = \frac{\Delta\mu/e}{\rho\lambda + R_I A} \approx \frac{I \Delta R_{NL}^0 \exp[d/(2\lambda_g)]}{\alpha(\rho\lambda + R_I A)}, \tag{3}$$

where $\Delta R_{NL}^0$ is the spin signal measured prior to depositing the island and the factor $\exp[d/(2\lambda_g)]$ is used to estimate the spin accumulation under the island from the injector-detector separation $d$ and the graphene spin diffusion length $\lambda_g = \sqrt{D\tau_s}$.

For each experiment, the best-fit results for $R_I A$ and the corresponding spin current $j_e$ are shown in Table 1. We find $R_I A = 3$ Ωμm$^2$, from which we calculate that the absorbed spin current is as large as $j_e = 10^8$ A/m$^2$. We discuss the physical significance of the extracted values of $R_I A$ in more detail below. Although one might be surprised that $\Delta R_{NL}$ decays quickly at small Fe thicknesses and then changes little with additional Fe thickness, this behavior is consistent with the assumption that $\lambda_{Fe} = 4$ nm. As the island is grown, its spin resistance decreases, but the absorbed spin current is ultimately limited by the series resistance of the interface. The local effective spin relaxation rate therefore increases with thickness and then saturates because of the completion of the graphene/Fe interface. The variation in $\Delta R_{NL}$ with thickness is explained by the completion of the interface at low thicknesses and the efficiency with which the spin absorption effect reduces the measured spin accumulation.

In Experiment 4, the process is repeated with a Cu island instead of Fe, where Cu is chosen because it has a longer spin diffusion length than Fe, is non-magnetic, has a weak chemical interaction with graphene, and has low spin-orbit coupling. We deposit the Cu island and monitor the spin signal $\Delta R_{NL}$ as a function of Cu thickness $t_{Cu}$. The result of Experiment 4, which is shown in Fig. 4(b), confirms that the decrease in $\Delta R_{NL}(t_{Cu})$ occurs over a larger thickness range than for Fe, as expected due to the longer spin diffusion length and larger spin resistance of Cu.

As before, the thickness dependence $\Delta R_{NL}(t_{Cu})$ is fit to the spin absorption model. We find that $R_I A$ = 4.8 $\Omega\mu m^2$. After depositing 80 Å of Cu, we calculate that $\Gamma_{Cu}$ = 418 ns$^{-1}$. In contrast, based on Experiment 1, the effective spin relaxation rate induced by Fe of the same thickness is three times larger than for Cu. We conclude that the spin absorption effect of Cu is smaller than Fe, as expected.

To determine the spin absorption by Fe across a nonmagnetic spacer, the deposition of Cu was halted after 80 Å and Fe deposition was started. The result is shown in Fig. 4(b), where the data are again fit to the spin absorption model. In this case, the total additional spin relaxation rate is $\Gamma = \Gamma_{Cu} + \Gamma_{Fe}(t_{Fe})$, where only the latter term increases with Fe thickness. We find a similar thickness dependence for Fe after Cu as compared to Fe directly on graphene, which is consistent with Fe absorbing a similar spin current in the graphene/Cu/Fe experiment as compared to the graphene/Fe experiments.

Table 1: Results of fitting the Fe thickness dependence of $\Delta R_{NL}$ to the spin absorption model. Experiment number, growth temperature, island width (as the percentage of the channel width covered by the island), charge injection current $I$ used for the measurement, product of the graphene resistance per square $R_{sq}$ and the diffusion constant $D$ fit from nonlocal Hanle data, best-fit interface resistance-area product $R_I A$, and spin current $j_e$ absorbed by the Fe in the limit $t_{Fe} \gg \lambda_{Fe}$ are indicated.

| Expt. | Temp (K) | Isl. Width (%) | $I$ ($\mu$A) | $R_{sq}D$ (k$\Omega$cm$^2$/s) | $R_I A$ ($\Omega\mu$m$^2$) | $j_e$ (A/m$^2$) into Fe |
|---|---|---|---|---|---|---|
| 1 | 20 | 30 | 1 | 39.1 | 2.8 | $1.5 \times 10^7$ |
| 2 | 20 | 85 | 5 | 146 | 3.1 | $1.0 \times 10^8$ |
| 3 | 300 | 90 | 1 | 95.6 | 12 | $2.4 \times 10^6$ |
| 4 (Fe on Cu) | 20 | 32 | 5 | 38.9 | gr/Cu: 4.8 Cu/Fe: 3.2 | $1.6 \times 10^6$ |

Finally, we discuss the physical significance of the extracted values for the interface resistance-area product. In the analysis presented above, the fitted values of the graphene/metal interface resistance-area product are approximately 10$^3$ times larger than either the Fe or Cu spin resistances. To understand the source of this interface resistance, we calculate the theoretical minimum resistance-area product $(R_I A)_{thy}$ of a graphene/metal interface from the number of available modes by $(R_I A)_{thy} = h/(4e^2 n)$, where $n$ is the 2-D graphene carrier concentration[29]. For $n = 10^{12}$ cm$^{-2}$, which is the approximate value at the gate voltages used, $(R_I A)_{thy} = 0.6$ $\Omega\mu$m$^2$. We conclude that the values for the interface resistance-area product that we measure are within an order of magnitude of the theoretical minimum possible value.

Given the low temperature growth of the island by molecular beam epitaxy, it is perhaps not surprising that the interfacial resistances are as small as those inferred from the model, with variations that reflect the degree of contamination before growth. We emphasize, however, that we cannot measure $R_I A$ values this small directly in the spin valve device geometry. It is therefore possible that the actual $R_I A$ product is *larger* than inferred from our model and that some other mechanism, such as proximity-induced magnetism[30-32] or enhanced spin-orbit coupling[33] at the graphene/metal interface, is leading to a larger interfacial spin relaxation rate. Our measurement cannot distinguish between interfacial spin relaxation in the presence of a larger $R_I A$ product and spin absorption. However, we emphasize that the experimental data can be interpreted purely in terms of spin absorption and that we observe no concrete evidence of interfacial spin relaxation. The model we introduce here places an upper bound on the maximum spin current absorbed by the island. The most practical option available for enhancing this value is increasing the spin accumulation in the channel, which will require further optimization of the injection contacts.

In conclusion, an *in situ* measurement of nonlocal spin transport is used to quantify the spin current absorbed by a small Fe island on a graphene surface. The Fe thickness dependence is interpreted using a 2D numerical simulation. We find that the data are consistent with a spin absorption model. Fitting the data to this model shows that the effective spin current absorbed by Fe can be as large as $10^8$ A/m² for an excitation current $I$ = 5 µA, and this absorbed spin current is limited by an interface resistance-area product of 3 Ωµm², which is nearing the theoretical minimum for a few-layer graphene/metal interface. A similar *in situ* study of a graphene/Cu/Fe junction is analyzed using the same model, where the effect of Cu is consistent with its longer spin diffusion length compared to Fe. Given the low resistance-area products achieved and the resulting bound on the absorbed spin current, these results suggest that new approaches will need to be considered for achieving the goal of non-local spin-transfer torque switching in few-layer graphene, for which effective spin current densities of order $10^{10}$ A/m² would be required.

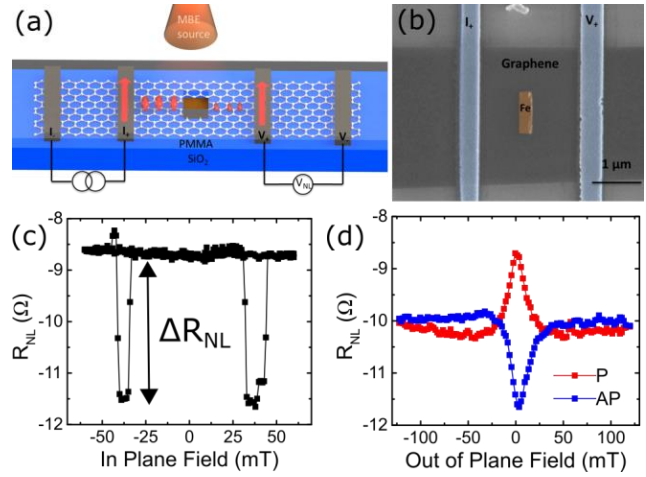

Figure 1: Graphene nonlocal spin valve island experiment. (a) Device layout and measurement configuration. (b) False color SEM image after measurement of the device used in Experiment 1. (c) Spin valve measurement taken prior to Fe deposition, indicating spin valve signal size $\Delta R_{NL}$. (d) Nonlocal Hanle measurement taken prior to Fe deposition, with parallel (P) and antiparallel (AP) injector and detector contact magnetization configurations.

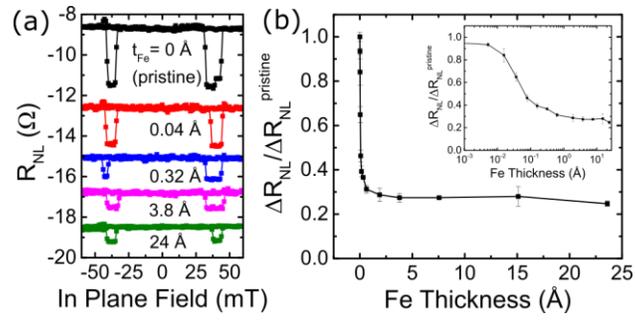

Figure 2: Effect of depositing the Fe island. (a) In-plane nonlocal magnetoresistance sweeps at different Fe thicknesses. Field sweeps are offset for clarity. (b) Normalized spin valve signal size as a function of Fe thickness, with inset showing the same data using a logarithmic scale for Fe thickness.

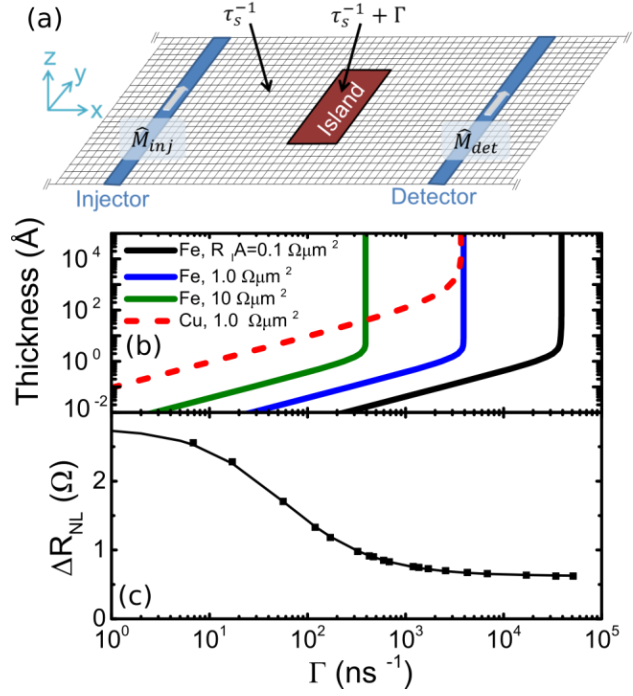

Figure 3: Analysis and modeling of the experiment. (a) Representation of the finite-element model, including additional spin relaxation rate $\Gamma$ under the island (not to scale). (b) Theoretical relationship between island thickness and $\Gamma$, for either an Fe or Cu island and various values of the graphene/metal interface resistance-area product $R_I A$. These theoretical curves assume spin diffusion lengths and monolayer thicknesses of $\lambda_{Fe} = 40$ Å and $t_{ML} = 1.43$ Å for Fe and $\lambda_{Cu} = 6000$ Å and $t_{ML} = 1.81$ Å for Cu. (c) Sensitivity of the spin valve signal size $\Delta R_{NL}$ to the additional spin relaxation rate for Experiment 1 determined by the finite-element model. Results of the model are shown as points and a smooth interpolation of these points is shown as a solid curve.

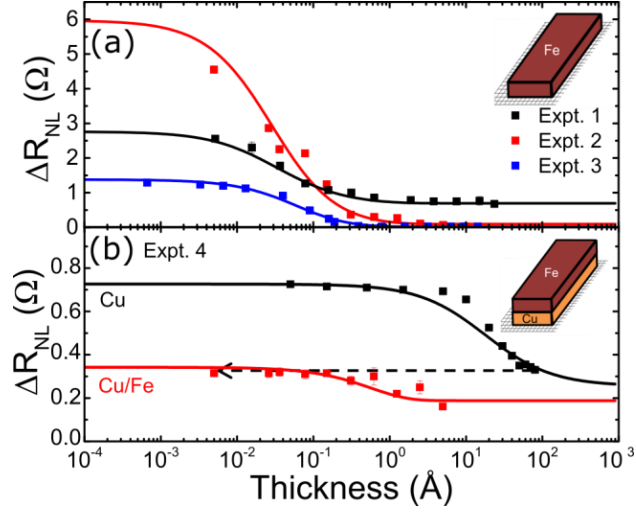

Figure 4: Fits to the spin absorption model. (a) Measurements of $\Delta R_{NL}(t_{Fe})$ for three different experiments, each fit to the spin absorption model. (b) Data and fits to the spin absorption model when Cu is deposited instead of Fe (black), and when Fe is deposited after depositing the Cu interlayer (red). The dashed black arrow indicates the measurement sequence. Results of the fits, including the graphene/metal interface resistance-area product and the calculated spin current, are shown in Table 1.

**Acknowledgments**

We thank Serol Turkyilmaz for his contributions to the development of the experiment.

[33] T. Frank, M. Gmitra, and J. Fabian, Theory of electronic and spin-orbit proximity effects in graphene on Cu(111), Phys. Rev. B **93**, 155142 (2016).

**Supplemental Material for "Spin absorption by *in situ* deposited nanoscale magnets on graphene spin valves"**

Table of contents:
**1)** Charge transport characteristics
**2)** Agreement between nonlocal Hanle data, analytical fit, and the finite-element model
**3)** Derivation of the theoretical thickness dependence $\Gamma(t)$ expected due to spin absorption
**4)** Finite-element modeling details
**5)** Additional details of each experiment

1) Charge transport characteristics

Prior to depositing the island, we measure the low bias contact resistance of each graphene/SrO/Co interface. We measure interface resistances of 1.8-85 k$\Omega$. In addition, we measure the gate-voltage dependence of the graphene resistance per square $R_{sq}$ as shown in Fig. S1.

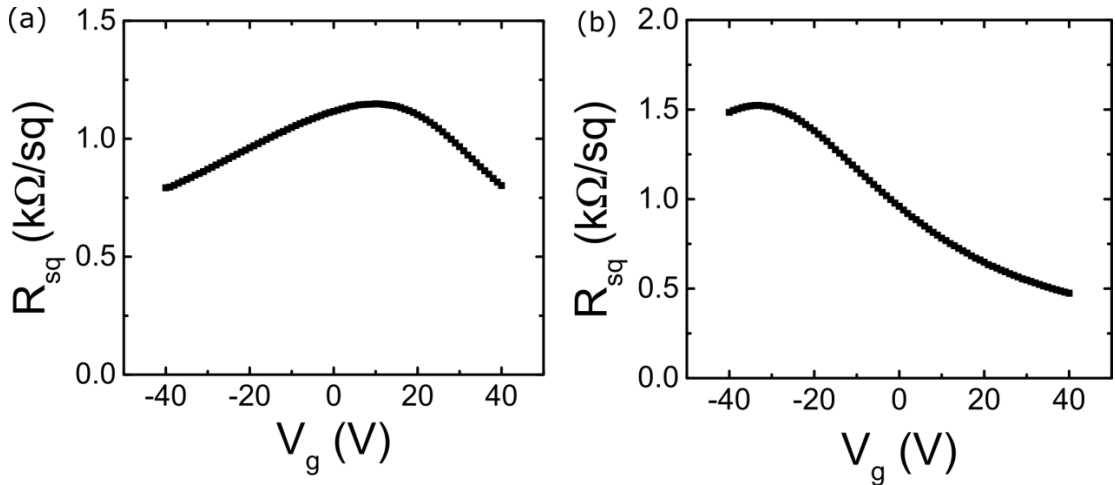

Figure S1: Gate-voltage dependence of the graphene resistance per square $R_{sq}$ for (a) Experiment 1 and (b) Experiment 4.

2) Agreement between nonlocal Hanle data, analytical fit, and the finite-element model

The nonlocal Hanle data taken prior to metal deposition are analytically fit to determine the spin lifetime, diffusion constant, and injection/detection efficiency as described in the text. The agreement between the data, analytical fit, and finite-element model for all four experiments is shown in Fig. S2.

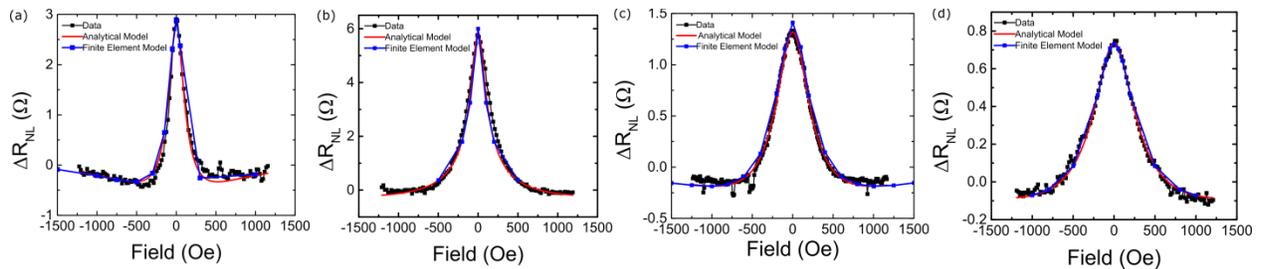

Figure S2: Comparison of the nonlocal Hanle data taken prior to island deposition to the analytical model and finite-element model for (a)-(d) Experiments 1-4, respectively.

3) Derivation of the theoretical thickness dependence $\Gamma(t)$ expected due to spin absorption

When a spin current is drawn out of the graphene, the local spin splitting of the chemical potential $\Delta\mu = \mu_\uparrow - \mu_\downarrow$ is reduced at rate $\Gamma$ such that $\Gamma\Delta\mu = (\partial\mu/\partial n)j_e/e$, where the thermodynamic inverse compressibility $\partial\mu/\partial n$ can be rewritten as $\partial\mu/\partial n = R_{sq}De^2$ by the Einstein relation. As in the main text, the actual spin current $j_s$ is scaled by $2e/\hbar$ to define an effective spin current $j_e$, $R_{sq}$ is the graphene square resistance and $D$ is the diffusion constant. In addition, the spin current $j_e$ is related to the spin resistance $R$ of the island by $j_e = \Delta\mu/(eR)$. Therefore, $\Gamma = R_{sq}D/R$. As derived elsewhere[1,2], the spin resistance of the island varies with island resistivity $\rho$ and thickness $t$ such that $R(t) = \rho\lambda/\tanh(t/\lambda)$. Finally, we assume that this absorbed spin current may be reduced by a series interface resistance $R_I$ that is constant once the interface (of area $A$) is completely formed, such that $R_IA(t) = R_IA/\tanh(t/t_{ML})$, where we assume the interface is fully formed after a monolayer of the island is deposited. Therefore,

$$\Gamma(t) = \frac{R_{sq}D}{\frac{\rho\lambda}{\tanh(t/\lambda)} + \frac{R_IA}{\tanh(t/t_{ML})}}, \tag{S1}$$

which is Eq. (2) in the main text. This model is described graphically in Fig. S3.

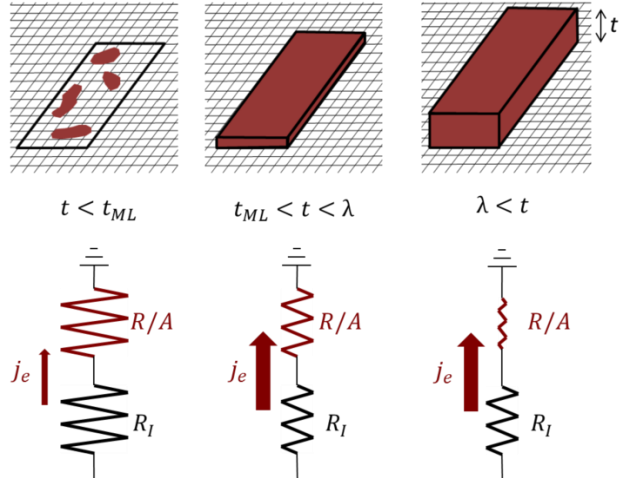

Figure S3: Diagrammatic representation of the effect of the island thickness $t$ on the island's spin resistance per unit area $R/A$ and the absorbed spin current $j_e$, including interface resistance-area product $R_IA$, relative to the monolayer thickness $t_{ML}$ and spin diffusion length $\lambda$ of the island.

Based on this spin absorption model and using $R_{sq}D$ known for a particular device, $\Gamma(t)$ can be theoretically calculated for given values of $\rho$, $\lambda$, $t_{ML}$, and $R_IA$. For example, the effect of the increased spin diffusion length and spin resistance of Cu compared to Fe is shown in Fig. S4. The assumptions for $\rho$, $\lambda$, and $t_{ML}$ are as used in Fig. 3(b) of the main text.

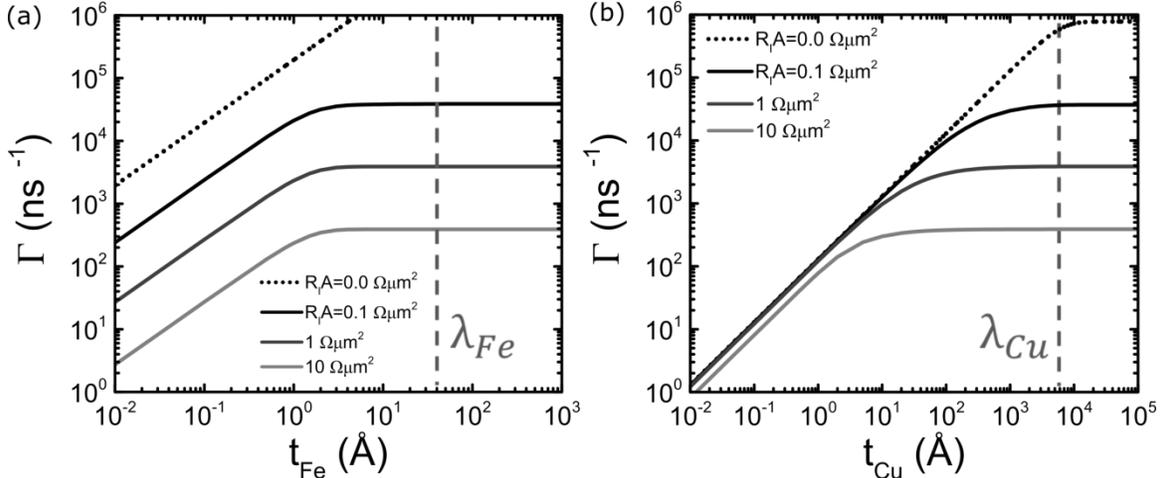

Figure S4: Contrast between the theoretical thickness dependence $\Gamma(t)$ for (a) Experiment 1, which uses Fe, and (b) Experiment 4, which uses Cu. In each case, different curves indicate the effect of varying the additional interface resistance-area product $R_I A$. The curves shown here assume $\lambda_{Fe} = 40$ Å, $\lambda_{Cu} = 6000$ Å, and $R_{sq}D = 39$ kΩcm²/s, where the latter matches both Experiments 1 and 4.

4) Finite-element modeling details

A 2-D environment composed of 25 nm x 25 nm cells is used to model the spin accumulation $\vec{S}(x,y)$ throughout the graphene channel, where $\vec{S}$ is the spin splitting of the chemical potential $\Delta\mu = \mu_\uparrow - \mu_\downarrow$ for spins oriented along each of the three cardinal directions ($i = x, y, z$). A forward Euler step algorithm with Neumann boundary conditions is used to evolve the distributed spin accumulation until steady state in accordance with spin diffusion, precession, and relaxation[3-4],

$$\frac{\partial \vec{S}}{\partial t} = D\nabla^2 \vec{S} - \gamma \vec{B} \times \vec{S} - \frac{\vec{S}}{\tau_s} + \dot{\vec{S}}_0(x,y) = 0, \tag{S2}$$

where $\gamma = 1.76 \times 10^{-2}$ Oe⁻¹ns⁻¹ is the gyromagnetic ratio, $\vec{B}$ is the applied magnetic field, and $\dot{\vec{S}}_0(x,y) = 2\alpha \widehat{M}_{inj} j_{inj} (\partial\mu/\partial n)/e$ is the rate of spin injection. The spin injection rate, which is nonzero only under the injector contact, is calculated from the fitted value of $\alpha$, the injected charge current density $j_{inj}$, and $\partial\mu/\partial n = R_{sq}De^2$. Spin absorption into the Co contacts is incorporated into the model as a local increase $\tau_c^{-1} = R_{sq}D/(R_C A)$ in the spin relaxation rate under the contacts, where $R_C$ is the measured contact resistance and $A$ is the contact area. The simulation is used to determine the component $S_y = \vec{S} \cdot \widehat{M}_{det}$ of the spin accumulation parallel to the detector magnetization, from which the spin valve signal size is calculated as $(\Delta R_{NL})_{model} = \alpha \langle S_y \rangle_{det}/I$, where $I$ is the injection charge current and $\langle S_y \rangle_{det}$ is the average of $S_y$ in the cells underneath the detector contact. The model is verified by confirming that the out-of-plane magnetic field dependence of $\Delta R_{NL}$ is consistent with the nonlocal Hanle data measured prior to metal deposition, as shown in Fig. S2.

We use this model to determine the sensitivity of $\Delta R_{NL}$ to additional spin relaxation in the graphene under the island. We do so by finding the steady-state solution to Eq. S2 for a range of values of $\Gamma$. Examples of $S_y(x,y)$ for various values of $\Gamma$ are shown in Fig. S5. As anticipated, we observe that the spin accumulation under the island is greatly reduced as $\Gamma$ increases. The effect of increasing $\Gamma$ is greater underneath the island than at the detector.

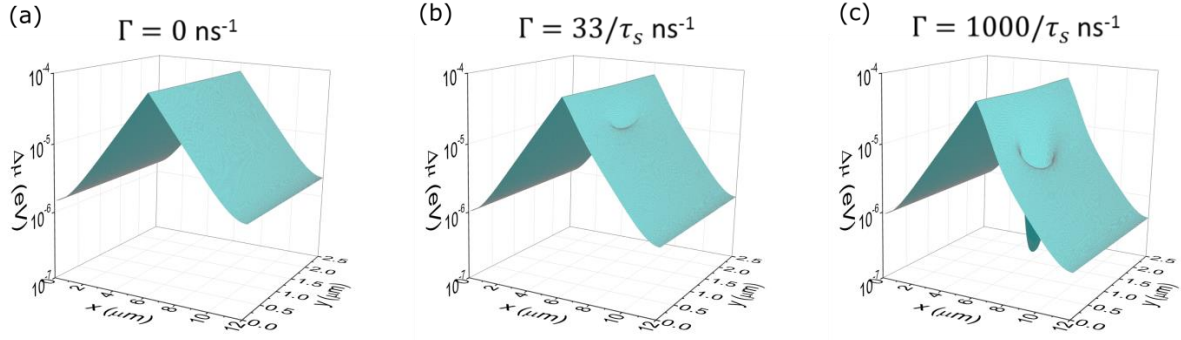

Figure S5: Steady-state solution to the finite element model corresponding to Experiment 1 for (a) $\Gamma = 0$, (b) $\Gamma = 33/\tau_s = 56$ ns$^{-1}$, and (c) $\Gamma = 1000/\tau_s = 1710$ ns$^{-1}$.

Finally, the sensitivity curves $\Delta R_{NL}(\Gamma)$ shown in Fig. S6 are calculated by finding the steady-state solution to Eq. S2 for many values of $\Gamma$ and interpolating between these results. This analysis is completed for each experiment, as the exact sensitivity $\Delta R_{NL}(\Gamma)$ depends on the geometry and spin transport parameters of each graphene channel.

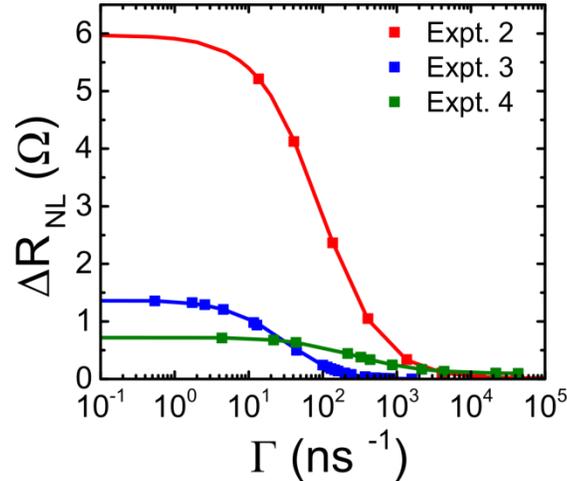

Figure S6: Sensitivity curves for Experiments 1-4 determined by simulation. These curves quantify how the introduction of additional spin relaxation rate $\Gamma$ under the island affects the spin signal $\Delta R_{NL}$.

5) Additional details of each experiment

The following table presents device information and the results of the analysis for all four experiments. The symbols $W$ and $d$ indicate the flake width and injector-detector separation, respectively. All values are calculated at the measurement temperature and $V_g = 0$ V prior to depositing the island.

| Experiment | 1 | 2 | 3 | 4 |
|---|---|---|---|---|
| Island metal(s) | Fe | Fe | Fe | Cu, Fe |
| $W$ (μm) | 2.5 | 2.7 | 2.0 | 1.9 |
| $d$ (μm) | 1.5 | 2.1 | 2.0 | 2.2 |
| Island length (nm) | 250 | 200 | 500 | 200 |
| $R_{sq}$ (Ω/sq) | 1120 | 393 | 956 | 194 |
| $\Delta R_{NL}^0$ (Ω) | 2.79 | 5.84 | 1.32 | 0.735 |
| $\tau_s$ (ps) | 585 | 736 | 310 | 230 |
| $D$ (cm$^2$/s) | 35 | 370 | 100 | 200 |
| $\alpha$ (%) | 11.3 | 11.3 | 8.0 | 10 |